\documentclass[twocolumn,showpacs,preprintnumbers,amsmath,amssymb]{revtex4}
\usepackage{graphicx}
\usepackage{dcolumn}
\usepackage{bm}

\begin{document}

\title{On the derivation of effective field theories}

\author{Dimo I. Uzunov} \altaffiliation[]{ Email address: d.i.uzunov@gmail.com.
Permanent address: Collective Phenomena Laboratory, Institute of
Solid State Physics, Bulgarian Academy of Sciences, BG-1784 Sofia,
Bulgaria.}

\affiliation{Department of Physics and Astronomy, University of
Western Ontario, 1151 Richmond St., London, Ontario, N6A 3K7,
Canada.}

\date{27 July 2008}% It is always \today, today,

\begin{abstract}
A general self-consistency approach allows a thorough treatment of
the corrections to the mean-field approximation (MFA). The natural
extension of standard MFA with the help of a cumulant expansion
leads to a new point of view on the effective field theories. The
proposed  approach can be used for a systematic treatment of
fluctuation effects of various length scales and, perhaps, for the
development of a new coarse graining procedure. We outline and
justify our method by some preliminary calculations. Results are
given for the critical temperature and the Landau parameters of
the $\phi^4$-theory -- the field counterpart of the Ising model.
An important unresolved problem of the modern theory of phase
transitions -- the problem for the calculation of the true
critical temperature, is considered within the framework of the
present approach. A comprehensive description of the ground state
properties of many-body systems is also demonstrated.

\end{abstract}



\maketitle

\section{\label{sec:level1}Introduction}

This investigation is focused on the correspondence between
microscopic models of phase transitions and their
quasi-macroscopic (field-theoretic) counterparts. Here we shall
outline a new self-consistency approach to a more accurate
derivation of effective field theories from microscopic models
defined on lattices.

Our method is general and can be used for a wide class of
microscopic models but for a concreteness here we shall illustrate
our approach with the Ising model (IM), given by
\begin{equation}
\label{eq1} {\cal{H}}(s) = -\frac{1}{2}\sum_{ij}^{N} J_{ij}s_is_j,
\end{equation}
where $s\equiv \{s_i\}$  denotes a lattice ``field'', $s_i=\pm 1$,
and the interaction constant of ferromagnetic type $J_{ij} =
J(|i-j|)>0$ depends on the intersite distance $|i-j|$ in a regular
$D$-dimensional lattice of $N$ sites (``spins'' or pseudo-spins).
Note, that $J(0)\equiv J_{ii} = 0$.

We shall follow the main path of the phase transition theory,
where the effective (quasi-macroscopic) field Hamiltonians (alias
Ginzburg-Landau (GL) free energies) are derived with the help of
two systematic methods: (i) Hubbard-Stratonovich transformations
(HST) and, (ii) a mean-field (MF) like
procedure~\cite{Uzunov:1993,Uzunov:1996, Uzunov1:1996}. Here we
propose a more thorough approach based on a convenient
generalization of (ii).

The known field theories exhibit both success and failure in the
description of many-body systems defined by microscopic models.
For example, we believe that the renormalisation group methods of
the modern theory of phase transitions~\cite{Uzunov:1993} based on
the $\phi^4$-theory yield a quite convenient description of the
scaling and universality properties of IM but we cannot be certain
that the field theory satisfactory describes important
non-universal properties, such as, for example, the critical
temperature $T_c$ and the lower critical dimensionality $D_L$ (for
IM, $D_L=1$, whereas within the $\phi^4$-theory, $D_L = 2$). In
fact the field theory fails along this line of studies. For
example, the fluctuation shift $(\Delta T_c)_f$ of $T_c$ predicted
within the one-loop approximation for the $\phi^4$-theory is a
very small and, hence, unrealistic, while in the higher orders of
the loop expansion this shift turns out infinite and its
calculation needs a special renormalization. As a result, the
problem for the value of $T_c$ within the framework of the present
field theory of phase transitions remains unresolved. Here we
shall show the genesis of this problem and present a satisfactory
solution.

The reason for the mentioned difficulties of the field theory
approach is in the quite simplified coarse-graining procedures
used in the derivation of the GL effective Hamiltonians from
microscopic models. HST is applied together with the
long-wavelength approximation (LWLA), namely, $(ka_0) \ll \pi$,
where $k=|\mbox{\boldmath$k$}|$ is the magnitude of the wave
vector $\mbox{\boldmath$k$} = (k_1,...,k_D)$,  $a_0$ is either the
lattice constant or, generally, the mean inter-particle distance.
LWLA leads to a correct expression for the Ornstein-Zernicke
correlation function but cuts the short-range (high-energy:
$\epsilon(k) \sim k^2, \Lambda < k <\phi/a_0$) interparticle
correlations of fluctuation type that have the main contribution
to the shift $(\Delta T_c)_f$; $\Lambda \ll \pi/a_0$ is the upper
cutoff for $k$ within LWLA.

Within the approach (ii) we usually say that we neglect the
fluctuations of the physical quantities from their equilibrium
values. This is not entirely true. It seems important to emphasize
that the latter are values in MF approximation (MFA) and, hence,
they are incorrect. Thus the fluctuations are defined towards
incorrect statistical averages, and their contribution to the free
energy of the system cannot be accepted as an entire fluctuation
effect. The latter can be correctly evaluated, if we are able to
define the fluctuations as variations towards exact statistical
averages calculated by the Hamiltonian ({\ref{eq1}). This task
seems unsolvable, but the present paper makes a step of
improvement of the theory along the same direction.

Note, that the methods (i) and (ii), no matter of the difference
between them, lead to the same GL effective field theory. Both
methods use  LWLA. For the method (ii) LWLA seems to me obligatory
because of the following important argument. HST can be applied
only to positively definite matrices $(J_{ij})$ but this condition
is not satisfied by  quite important lattice models, such as the
nearest-neighbour ($nn$) IM. In  LWLA, however, the same
interaction matrix $J_{ij;nn}$ of IM is modified to a form that is
positively definite. Besides, in both methods,  LWLA is used to
help the derivation of the quasi-macroscopic (fluctuation) and
macroscopic (thermodynamic) properties, i.e., the LWLA is a tool
of a ``coarse-graining'' procedure for the many-body system. But
in both cases (i) and (ii), the micro- and mesoscopic
interparticle correlations are ignored. Here we shall use a
generalization of the method (ii) in order to improve this
disadvantage of the theory. In this way we shall present more
accurate calculation of the LG parameters (vertices) of the
effective field theory of Ising systems. Besides we shall show
explicitly for the first time the mechanism of ``statistical
correlation'' which leads from the two-site ($i-j$) trivial
correlation presented by the initial interaction $J_{ij}$ to
short-, meso-, and large-scale effective multiparticle
correlations of fluctuation type -- fluctuation correlations. Thus
we shall establish and develop for the first time a new
``coarse-graining'' procedure, and this is the main aim of our
report. Besides, we shall show that the GL parameters of the
$\phi^4$-theory acquire ($1/z$)-corrections for both short-range
and long-range interactions $J(|i-j|)$~\cite{Uzunov:1993}. These
corrections will be presented to second order in $(1/z)^2$.

The $1/z$-expansion has been introduced by R.
Brout~\cite{Brout:1959} and applied in calculations of $T_c$ and
thermodynamic susceptibilities~\cite{Horwitz:1961, Englert:1963,
Stinchcombe:1963, Fisher:1964, Brout:1965, Vaks:1966, Vaks1:1967,
Vaks2:1967, Brout:1974}. We are not aware of another relevant work
along this line of research except for our recent
investigation~\cite{Uzunov:1996, Uzunov1:1996}}; the latter will
be used in our investigation. Remember, that the (Brout) approach
is a development of an older method -- the Kirkwood method of
cumulants (semi-invariants)~\cite{Kirkwood:1938}; hence, we shall
follow a cumulant expansion which is well known.

We shall  generalize the Brout approach in a way that makes
possible to derive effective field theories, and this will allow
us to reveal new and surprising features of many-body systems. In
the Brout scheme, the mean (``molecular'') field is spatially
($i$-) independent. We find that there are no physical reasons for
this assumption and extend the ``mean-field concept.'' Within our
approach, the so-called mean field is spatially dependent up to
the moment when one should find the actual ground state. Then the
fluctuation phenomena occur with respect to this ground state. In
our approach the GL parameters and, hence, the ground state have a
more precise evaluation.

Our consideration has been performed for interactions $J(|i-j|)$
of a quite general type, namely for all interactions that can be
presented by the equality
\begin{equation}
\label{eq2} \sum_j J_{ij} = zJ_0 \equiv J,
\end{equation}
where $J_0$ is an effective exchange constant, and $z$ is an
effective ``coordination'' number (number of interacting
neighbors). In particular cases our results will be referred to
the most common case of $nn$ interactions (then $n=2D$ for simple
lattices). For a simplicity, here we shall assume that IM is
defined by a simple cubic ($sc$) lattice. The consideration can be
easily expended to other types of regular lattices, as well as to
irregular lattices with certain forms of quenched disorder, for
example, random potential ~\cite{Uzunov:1993}. We introduce the
interaction radius by $R_{int} \approx z^{1/D}$ -- a quantity,
which is equal to the so-called zero-temperature correlation
length (see Section 2). Let  emphasize that our results can be
rederived without difficulties in the presence of an external
field $\{h_i\}$ conjugate to the lattice field $\{s_i\}$.

Other theories intended to improve the coarse graining procedures
of many-body systems are, for example, the hierarchical reference
theory~\cite{Reatto:1995} and a method of collective
variables~\cite{Yukhnovskii:1980}. The former is based on a
particular renormalization procedure and yields interesting
results for the effective critical exponents of continuous phase
transitions. The latter theory uses a variant of cumulant
expansion and collective variables which, in certain limit, form
the order parameter field. In these features the collective
variable theory~\cite{Yukhnovskii:1980} resembles our approach
but, generally, in contrast to the latter, it significantly
departs from the original Brout idea. The mentioned theories does
not use $1/z$-expansion and for this reason and apart from some
common aims and general ideas, cannot easily be compared with the
present approach, which is a very direct extension of the
many-body theory beyond the standard mean-field (``tree")
approximation.

In Sec. II we present a general approach to the treatment of
fluctuation correlations at various length scales and discuss
aspects of the usual theory that corresponds to the lowest order
approximation of a perturbation expansion of cumulant type. In
Sec. III we investigate the higher orders of the mentioned
perturbation expansion and demonstrate several new features of the
effective field theory. Our main results are summarized and
discussed in Sec. III.D -- III.G.

\section{\label{sec:level1}Present status of the effective field theory}

\subsection{General scheme}

The equilibrium  free energy of IM as
a function of the temperature $T$ and the field configuration $h =
\{h_i\}$ is given by
\begin{equation}
\label{eq3} G(T) = -\beta^{-1}\mbox{ln}\left\{\mbox{Tr}e^{-\beta
{\cal{H}}(s)}\right\},
\end{equation}
where $\beta^{-1} = k_BT$, and the Trace is over the allowed
lattice configurations $s = \{s_i\}$. Note, that the equilibrium
values of the physical quantities are calculated as averages with
respect to the statistical ensemble based on the Hamiltonian
(\ref{eq1}) and, in particular, the statistical averages of type
$\langle s_i...s_j\rangle$ are obtained as derivatives of the
partition sum whereas the irreducible averages $\langle\langle
s_i...s_j\rangle\rangle =\langle(s_i-\langle
s_i\rangle)...(s_j-\langle s_j\rangle)\rangle$ are obtained as
derivatives of the Gibbs free energy (\ref{eq3}).

Let us introduce the shift
\begin{equation}
\label{eq4} s_i = \phi_i + \delta s_i,
\end{equation}
where the lattice field $\phi_i$ is an arbitrary (auxiliary) field
configuration that is not necessarily associated with the averaged
spin $\langle s_i\rangle$ at site $i$, and the ``fluctuation''
$\delta s_i$ is merely the difference $(s_i - \phi_i)$. The
identification of $\phi_i$ with a statistical average
$\langle...\rangle$ over the full Hamiltonian~(\ref{eq1}), or,
with a statistical average $\langle...\rangle_0$ corresponding to
another ensemble as well as the interpretation of $\delta s_i$ as
a fluctuation around $\langle s _i\rangle$ (or $\langle
s_i\rangle_0$) may be a matter of further considerations. At this
stage $\phi_i$ and $\delta s_i$ are auxiliary variables which obey
(\ref{eq4}) and are not referred to concrete physical quantities.

Following a standard procedure (see, e.g.,
Refs.~\cite{Uzunov:1993, Uzunov:1996}) we obtain the following
effective non-equilibrium free energy
\begin{equation}
\label{eq5} {\cal{H}}(\phi) =  {\cal{H}}_0(\phi) +
{\cal{H}}_f(\phi),
\end{equation}
with $\phi \equiv \{\phi_i \}$,
\begin{equation}
\label{eq6} {\cal{H}}_0(\phi) =
\frac{1}{2}\sum_{ij}J_{ij}\phi_i\phi_j - \beta^{-1}\sum_i\mbox{ln}
\left[2\mbox{ch}(\beta a_i)\right].
\end{equation}
Here
\begin{equation}
\label{eq7} a_i = \sum_j J_{ij}\phi_j
\end{equation}
is the ``mean'' (molecular) field, and
\begin{equation}
\label{eq8} {\cal{H}}_f(\phi) = -\beta^{-1}\mbox{ln}\left<
\mbox{exp}\left[\frac{\beta}{2}\sum_{ij}J_{ij}(s_i-\phi_i)(s_j-\phi_j)\right]\right>_0
\end{equation}
is the ``fluctuation'' part. As usual, we shall often call the
free energy~(\ref{eq5}) an ``effective Hamiltonian'' (see, also,
Ref.~\cite{Note:2008}).

In~(\ref{eq8}), $\langle...\rangle_0$ denotes a statistical
average over an ensemble defined by the auxiliary (``MF'')
Hamiltonian
\begin{equation}
\label{eq9} {\cal{H}}_a (\phi,s) = - \sum_i a_i s_i;
\end{equation}
the respective partition function and (nonequilibrium) free energy
are given by ${\cal{Z}}_a(\phi) = \mbox{Tr}\left[\mbox{exp}(-\beta
{\cal{H}}_a)\right]$, and $G_a(\phi) =
-\beta^{-1}\mbox{ln}{\cal{Z}}_a(\phi)$. For this simple ensemble,
we have
\begin{equation}
\label{eq10} \langle s_i \rangle_0 = \mbox{th}\left[\beta
a_i(\phi)\right].
\end{equation}
The calculation of averages of  type $\langle s_i...s_j\rangle_0$
as well as ``irreducible'' averages of type $\langle\langle \delta
s_i...\delta s_j\rangle\rangle_0$ is also straightforward. These
averages represent a form of fluctuation correlations, but they
are just an auxiliary theoretical tool rather than real objects.
In contrast, the real objects, namely, (full) statistical averages
$\langle...\rangle$ and $\langle\langle...\rangle\rangle$ within
the total Hamiltonian (\ref{eq1}) cannot be exactly calculated.

\subsection{Usual theory}

In the framework of the usual theory, the ``fluctuation'' term
${\cal{H}}_f$ is ignored. This is the MFA. In the present general
format of the theory, we have $N$ (self-consistency) equations of
state ($\partial {\cal{H}}/\partial \phi_i) = 0$ at fixed $T$ (and
$h=\{h_i\})=0)$ -- one equation per a lattice vertex $i$:
\begin{equation}
\label{eq11} \sum_j J_{ij}\left\{\bar{\phi}_j -
\mbox{th}\left[\beta a_j(\bar{\phi})\right]\right\} =0,
\end{equation}
where $\bar{\phi} = \{\phi_i\}$. It is easy to see that the number
($z-1$) of nonzero terms ($i\neq j$) in all $N$ sums~(\ref{eq11})
is equal to the number of nonzero interaction constants acting on
the site $i$: $J(|i-j|) > 0$ for $a_0 \leq |i-j| \leq R_{int}$;
$J(|i-j|=0$ for $|i-j| \equiv R > R_{int}$. The ``equations of
state'' (\ref{eq11})  can be written in the simple form
\begin{equation}
\label{eq12} \bar{\phi}_i = \mbox{th}\left[\beta
a_i(\bar{\phi})\right].
\end{equation}
The equivalence of~(\ref{eq11}) and (\ref{eq12}) can be easily
proven for any number $N \geq 1$.

From (\ref{eq10}) and (\ref{eq12}) we obtain
\begin{equation}
\label{eq13} \bar{\langle s_i \rangle}_0 = \mbox{th}\left[\beta
a_i(\bar{\phi})\right] = \bar{\phi}_i.
\end{equation}

Thus in this quite general form of MFA (${\cal{H}}_f \approx 0$),
we have: $\langle s_i \rangle_0 = \bar{\phi}_i$,
$\langle...\rangle = \langle...\rangle_0$, and $G(T,h) =
{\cal{H}}_0(\bar{\phi})$ is the equilibrium free energy that
corresponds to extrema (including minima) $\bar{\phi}$ of
${\cal{H}}(\phi) \approx {\cal{H}}_0(\phi)$ -- the non-equilibrium
MF free energy given by the non-equilibrium (arbitrary) order
parameter field $\phi_i$. By $\bar{\phi_i}$ from~(\ref{eq11}) we
denote the equilibrium configuration of the latter; hereafter the
``bar'' of the equilibrium value $\bar{\phi}_i$ of $\phi_i$ will
be often omitted. Now one may perform a Landau expansion for small
$\bar{\phi}_i$ in order to get other known forms of the MF theory.

However, at the present stage of consideration we are interested
in some more generality and for this reason we continue our
discussion of the non-equilibrium free energy functional
\begin{equation}
\label{eq14} \tilde{G}(T/\phi) \equiv {\cal{H}}(T/\phi) \approx
{\cal{H}}_0(T/\phi)
\end{equation}
as given by Eq.~(\ref{eq6}). In (\ref{eq14}), the dependence of
the non-equilibrium free energy $\tilde{G}$ on $\phi$ is denoted
by ``$/\phi$'' because of the more special role of this variable,
namely, the variation $\delta \tilde{G}$ should be zero at thermal
equilibrium and from this condition one obtains the possible
thermal equilibria $\bar{\phi}$ (alias ``self-consistency
condition'').

The expansion of the $log-$term in~(\ref{eq6}) up to fourth order
in $\phi_i$ yields the known result for the lattice
$\phi_i^4-$theory of the Ising model:
\begin{widetext}
\begin{equation}
\label{eq15} {\cal{H}}_0(\phi) =
\frac{1}{2}\sum_{ij}J_{ij}\phi_i\phi_j
-\frac{\beta}{2}\sum_{ijk}J_{ij}J_{ik}\phi_j\phi_k +
\frac{\beta^3}{12}\sum_{ijklm}J_{ij}J_{ik}J_{il}J_{im}\phi_j\phi_k\phi_l\phi_m.
\end{equation}
\end{widetext}
One may apply  LWLA to this form of the theory but we shall follow
a different path.

\subsection{Continuum limit}

Using the rule
\begin{equation}
\label{eq16} \sum f_i = \rho \int d^Dx f(\mbox{\boldmath$x$})
\equiv \rho \int
 d\mbox{\boldmath$x$} f(\mbox{\boldmath$x$})\:,
\end{equation}
where $\rho = (N/V)$ we can write Eq.~(\ref{eq2}) in the form
\begin{equation}
\label{eq17} J = \rho \int d^DR J(R),
\end{equation}
with $R = |\mbox{\boldmath$R$}|$; $\mbox{\boldmath$R$} =
(\mbox{\boldmath$x$} -\mbox{\boldmath$y$})$. Note, that the factor
$\rho$ in (\ref{eq19}) - (\ref{eq21}) can be avoided and this is
the usual practice. In the latter case the physical dimension of
the respective physical quantity is changed by a factor $[V] \sim
[L]^D$; for example,  $[L]^D[f_i] = [f(\mbox{\boldmath$x$})]$. Of
course, one may use both variants.

In the continuum limit the effective Hamiltonian (\ref{eq6})
corresponding to a zero external field ($h=0$) takes the form
\begin{eqnarray}
\label{eq18} {\cal{H}}_0 & = &\frac{\rho}{2} \int
d^Dx\phi(\mbox{\boldmath$x$})I[\phi(\mbox{\boldmath$x$})] \\
\nonumber &&  - \rho\beta^{-1}\int d^Dx
\mbox{ln}\:\mbox{ch}\left\{\beta^{-1}I\left[\phi(\mbox{\boldmath$x$})\right]\right\},
\end{eqnarray}
where
\begin{equation}
\label{eq19} I\left[ \phi(\mbox{\boldmath$x$})\right] = \rho \int
d^DyJ(R)\phi(\mbox{\boldmath$y$})\:.
\end{equation}

Now we apply LWLA in the form
\begin{equation}
\label{eq20} \left[\phi(\mbox{\boldmath$x$}) -
\phi(\mbox{\boldmath$y$})\right]^2 \ll
|\phi(\mbox{\boldmath$x$})\phi(\mbox{\boldmath$y$})|
\end{equation}
and under this assumption truncate the Taylor expansion
\begin{eqnarray}
\label{eq21} \phi(\mbox{\boldmath$y$})& =&
\phi(\mbox{\boldmath$x$}) + \sum^{D}_{\alpha = 1}
\frac{\partial\phi(\mbox{\boldmath$x$})}{\partial
x_{\alpha}}R_{\alpha} \\ \nonumber && +
  \frac{1}{2}\sum^D_{\alpha,\beta
    =1}\frac{\partial^2\phi(\mbox{\boldmath$x$})}{\partial x_{\alpha}\partial
    x_{\beta}} R_{\alpha}R_{\beta} + ...
\end{eqnarray}
to the second order in $\mbox{\boldmath$R$} =
\left\{R_{\alpha}\right\}$.

Using the approximation (\ref{eq20}), (\ref{eq19}) becomes
\begin{equation}
\label{eq22} I\left[\phi(\mbox{\boldmath$x$})\right] =
J\phi(\mbox{\boldmath$x$}) + \frac{\tilde{J}}{2D}
\nabla^2\phi(\mbox{\boldmath$x$}),
 \end{equation}
where $J$ is given by (\ref{eq17}), and
\begin{equation}
\label{eq23} \tilde{J} = \rho \int d^DRJ(R)R^2.
\end{equation}
Now we substitute (\ref{eq22}) in (18), perform the expansion up
to  order $\phi^4(\mbox{\boldmath$x$})$ and to second order in
$\nabla \phi(\mbox{\boldmath$x$})$. Besides, we should keep in
mind, that in expansion in powers of $\phi_i$ we cannot
distinguish between $T$ and $T_{c0}$ except for the
$\phi^2_i-$term where the difference between $T$ and $T_{c0}$
should be kept only to the lowest nonvanishing order; in our case,
this is the first order in $(T-T_{c0})$: see, e.g.,
Ref.~\cite{Uzunov:1993}. Following these notes, we perform at a
certain stage of the calculation an integration by parts with the
convenient boundary condition $\nabla \phi(\mbox{\boldmath$x$}) =
0$ and  obtain the well known GL effective Hamiltonian
\begin{equation}
\label{eq24} {\cal{H}}_0 = \rho\int d^D x
\left\{\frac{\tilde{c}_0}{2}\left[\nabla\phi(\mbox{\boldmath$x$})\right]^2
   + \frac{r_0}{2}\phi^2(\mbox{\boldmath$x$}) +
u_0\phi^4(\mbox{\boldmath$x$})\right\},
\end{equation}
with
\begin{equation}
\label{eq25} c_0 = \frac{R^2_{int}}{2D}J,\;\;\;  r_0 (T) =
k_B(T-T_{c0}),\;\;\; u_0 = \frac{J}{12}.
\end{equation}
Here $T_{c0} = (J/k_B)$ and terms of order $t_0 =
(T-T_{c0})/T_{c0} \ll 1$ have been neglected in $c_0$ and
$u_0$~\cite{Uzunov:1996}, i.e. these two parameters are calculated
at $T_{c0}$~\cite{Uzunov:1993, Uzunov:1996}.

To clarify the result (\ref{eq24}) - (\ref{eq25}) we shall mention
that $J(R)$ for $R > R_{int}$ is very small and can be ignored.
Setting $J(R) \sim J_0$ in (\ref{eq17}), comparing the result with
$J=zJ_0$ from (\ref{eq2}), and noticing that $\rho = 1/v \sim
a_0^D$,  one obtains $z \sim \left(R_{int}/a_0\right)^D$ as should
be, and $J \approx J_0 R^D_{int}$. In the same way one gets
$\tilde{J} \approx J R^2_{int}$.

The results (\ref{eq24}) - (\ref{eq25}) show that the energy of
the
 spatially dependent configurations of the field depends on the interaction
 radius. The latter serves as a coherence (correlation) length of the field
$\phi(\mbox{\boldmath$x$})$; see the parameter $c_0$ given in
(\ref{eq25}).
 In order to clarify this point, let us consider the so-called zero-temperature
 correlation length~\cite{Uzunov:1993}, defined by
$\xi_0 \equiv \xi(T=0) = \left[-c_0/r_0(0)\right]^{1/2}$. Using
(\ref{eq25}) and $T_{c0} = J/k_B$ we obtain  $\xi_0 =
R_{int}\sqrt{2D}$.

 \subsection{Discussion}

 Note that the temperature range of validity of these
considerations is $t_0(T) \ll 1$ and $(ka_0) \ll \pi$. The
spatially dependent fluctuations correspond to a higher energy
than the uniform configuration, and hence the latter contains the
deepest (global) minima $\bar{\phi}$ of the effective free energy
nevertheless we have written N ``equations of state'' as a result
of the minimization of the effective Hamiltonian. Now one can
easily show that the variation of the Hamiltonian (\ref{eq24})
with respect to the field will give again spatially dependent
solutions but in the usual theory they are interpreted as
``spatially dependent fluctuations'' which, together with the
uniform fluctuation $\delta \phi = (\phi -\bar{\phi})$ towards the
stable state $\bar{\phi}$, are all fluctuations in the system in
LWLA. But we know that this picture contains the approximation
$\langle s_i\rangle \sim \langle s_i\rangle_0 = \bar{\phi}$.

Our point of view is the following. The averages
$\langle\rangle_0$ should not be taken very seriously. They are an
auxiliary tool in our consideration, and are not the final aim of
our investigation. We have used these averages only because they
appear along our way of obtaining an effective Hamiltonian
${\cal{H}}$ (or ${\cal{H}}_0$ in the lowest order of the theory)
in which the statistical degree of freedom  $\phi_i$  varies in a
wide range of values $(-\infty < \phi_i < \infty)$. This is the
only consistent interpretation of our consideration performed so
far. We can assume that up to now we have obtained nothing else
but an effective free energy (effective lattice Hamiltonian) in
terms of the new lattice field $\phi_i$. In a clear approximation,
this effective model is given by~(\ref{eq6}); for the expansion in
powers of $\phi_i$, see~(\ref{eq15}).

Another important aspect of our consideration is that the new
Hamiltonian contains more $ij-$ interactions than the original
 model~(\ref{eq1}). The mathematical form of (\ref{eq15})
gives a clear physical interpretation of these interactions: the
first $\phi_i\phi_j-$term in the r.h.s. of ~(\ref{eq15}) describes
an interaction that is quite similar to the original inter-spin
interaction in~(\ref{eq1}) whereas the second term of the same
type in~(\ref{eq15}) describes an indirect two-site
$jk-$interaction that is mediated by the ``$i-$spins''. This means
that the latter interaction $J_{ij}J_{ik}$ has twice larger radius
of action than the original $J_{ij}$ exchange. The four-point
interaction given by the third term in the r.h.s. of~(\ref{eq15})
can also be described in the above style, introduced for the first
time in this paper.

Therefore, even at this early stage of consideration
(${\cal{H}}\approx {\cal{H}}_0$) we see that the effective free
energy exhibits effects of ``statistical extension of the original
inter-particle correlations (interactions)''. This ``principle of
growth of statistical correlations'' is well known in the general
phase transition theory. Here we show the concrete mechanism of
the respective phenomenon and a systematic way of description of
growth of statistical correlations. This point will become more
clear from the results in the next Section. We shall see that the
investigation along this path leads to a quite unexpected and
intriguing picture.

\section{\label{sec:level1} Beyond the standard theory}

\subsection{Perturbation series}

 Here we consider the $\phi-$contributions
to the effective free energy (Hamiltonian) ${\cal{H}}(\phi)$ which
are generated by the term ${\cal{H}}_f(\phi)$.  For obvious
reasons, terms that are $\phi_i$-independent will be omitted.

It seems convenient to rearrange our theory by introducing the
auxiliary variables $\Delta_i = (s_{i0} - \phi_i)$, and $\sigma_i
= (s_i-s_{i0})$, where $s_{i0} \equiv \langle s_i \rangle_0$. Then
${\cal{H}}$ can be written as an infinite perturbation series in
powers of the (perturbation) Hamiltonian part
\begin{equation}
\label{eq26} S_f(s,\phi) =
-\frac{1}{2}\sum_{ij}J_{ij}\sigma_i\sigma_j -
\sum_{ij}\Delta_i\sigma_j.
\end{equation}
The respective series can be presented in the form
\begin{equation}
\label{eq27} {\cal{H}}_f(\phi) =
-\frac{1}{2}\sum_{ij}J_{ij}\Delta_i\Delta_j +
\sum_{l=1}^{\infty}{\cal{H}}_f^{(l)}(\phi),
\end{equation}
where
\begin{equation}
\label{eq28} {\cal{H}}_f^{(l)}(\phi) =
\frac{(-\beta)^{l-1}}{l!}\langle S_f^l(s,\phi)\rangle_{0c},
\end{equation}
$\langle...\rangle_{0c}$ denotes the so-called connected
averages~\cite{Uzunov:1993}; for example, the average $\langle
S_f^2 \rangle_0\langle S_f^2\rangle_0$ is excluded from the
connected $\langle S_f^4\rangle_{0c}$. For this cumulant
(semi-invariant) expansion rules, similar to the Wick theorem in
the perturbation theory of propagator type, are not available and
one should perform the calculations with some caution. The term
${\cal{H}}_f^{(1)}$ is equal to zero, and this leads to a
reduction of some infinite series in the next orders of the theory
$(l > 1)$.

\subsection{Lowest order correction}

Let us consider the first term in the r.h.s. of Eq.~(\ref{eq27})
and neglect all others. This yields ${\cal{H}} = ({\cal{H}}_0 +
{\cal{H}}_f)$ in the form
\begin{widetext}
\begin{eqnarray}
\label{eq29} {\cal{H}}& \approx &
\frac{\beta}{2}\sum_{ijk}J_{ij}J_{ik}\phi_j\phi_k -
\frac{\beta^2}{2}\sum_{ijkl}J_{ij}J_{ik}J_{jl}\phi_k\phi_l  \\
\nonumber && -\frac{\beta^3}{4}\sum_{ijklm}J_{ij}J_{ik}J_{il}
J_{im}\phi_j\phi_k\phi_l\phi_m
+\frac{\beta^4}{3}\sum_{ijklmn}J_{ij}J_{jk}
J_{il}J_{im}J_{in}\phi_k\phi_l\phi_m\phi_n.
\end{eqnarray}
\end{widetext}
 Performing this straightforward
calculation one readily sees a very important property of the
present theory, namely, that the first term in the r.h.s.
of~(\ref{eq15}) is totally compensated by a respective counter
term coming from the new contribution ($\sim \Delta_i\Delta_j$) to
the effective free energy. Besides, another new term {\em twice}
compensates the second term in~(\ref{eq15}) so that the term of
type $JJ\phi\phi$ now appears with a positive sign. The $\phi^4-$
part of the effective free energy also undergoes a drastic change
due to the $\Delta\Delta-$correction coming from Eq.~(\ref{eq28}).

The same result can be obtained in a more general and, perhaps,
more convenient way, if we add the $\Delta\Delta-$ term
in~(\ref{eq27}) to ${\cal{H}}_0$ from~(\ref{eq6}) before doing
 the expansion of Landau type. Then, within the
same lowest order approximation for the series~(\ref{eq27}) we
obtain a more general result for ${\cal{H}}$, namely
\begin{eqnarray}
\label{eq30} {\cal{H}}& = & -
\frac{1}{2}\sum_{ij}J_{ij}\mbox{th}(\beta a_i)\mbox{th}(\beta
a_j))
   \\ \nonumber && +\sum_{ij}J_{ij}\phi_i\mbox{th}(\beta a_j)
 - \beta^{-1}\sum_i\mbox{ln} \left[2\mbox{ch}(\beta a_i)\right].
\end{eqnarray}
This form of ${\cal{H}}$ clearly shows the lack of the simple
$J_{ij}\phi_i\phi_j$ term describing the direct two-site exchange.
In our further considerations we shall be faced only with
inter-particle interactions (correlations), which are extended at
distances larger than $R_{int}$. In order to obtain~(\ref{eq29})
one must expand the transcendental functions in~(\ref{eq30});
$(\beta a_i) \ll 1$.

The new forms (\ref{eq29}) and (\ref{eq30}) of the effective free
energy describes only indirect two-site interactions because the
direct two-site interaction disappeared from our consideration.
Now we are at a stage of description of correlations which extend
up to $2R_{ij}$ and larger distances. The tendency of the growing
length scale of the interactions included in our effective free
energy will be the main and unavoidable feature of our further
consideration.

Another very important feature of our findings is that in the
simplest variant of the theory when the field is uniform
$(\phi_i\equiv\phi)$ as well as in the GL variant in LWLA given by
Eq.~(\ref{eq24}) the values of the Landau coefficients $c_0$,
$r_0$, and $u_0$ do not change within the framework of accuracy of
the effective $\phi^4$-field theory, where terms (Landau
invariants) of order $O(t^3)$ are ignored; $t = (T-T_c)/T_c$. This
property seems to exist to any order of the expansion~(\ref{eq27})
in the limit of an infinite-range ($z \rightarrow \infty$) initial
interaction $J_{ij}$.  The $1/z-$ corrections to the parameters
($c_0,r_0,u_0$) of the effective field theory are obtained from
the $(l>1)-$terms in the series~(\ref{eq27}).

\subsection{High-energy corrections of higher order}

We have already mentioned that the $(l=1)$-term in~(\ref{eq27}) is
zero. A calculation of the next two terms~($l=2,3$) in
(\ref{eq27}) has been carried out in Ref.~\cite{Uzunov:1996}.
Following Ref.~\cite{Uzunov:1996} we can write ${\cal{H}}_f^{(2)}$
in the form
\begin{eqnarray}
\label{eq31} {\cal{H}}^{(2)}_f & = & -
\frac{\beta}{4}\sum_{ij}\frac{J^2_{ij}}{\mbox{ch}^2(\beta
a_i)\mbox{ch}^2(\beta a_j)} \\ \nonumber &&
-\frac{\beta}{2}\sum_{ijk}J_{ij} J_{ik}
\frac{\Delta_j\Delta_k}{\mbox{ch}^2(\beta a_i)}.
\end{eqnarray}
The result~(\ref{eq31}) gives the first $(1/z)-$corrections to the
parameters $c_0, r_0$, and  $u_0$. Adding the result (\ref{eq31})
to ${\cal{H}}(\phi)$ we obtain a form of the effective Hamiltonian
${\cal{H}}(\phi)$ which is more precise than the preceding ones.
Let us write down this quite lengthy expression of the effective
Hamiltonian:
\begin{widetext}
\begin{eqnarray}
\label{eq32} {\cal{H}}& = &
\frac{\beta^2}{2}\sum_{ijkl}J_{ij}J_{ik}J_{jl}\phi_k\phi_l +
\frac{\beta^3}{2}\sum_{ijkl}J^2_{ij}J_{ik}J_{il}\phi_k\phi_l
-\frac{\beta^3}{2}\sum_{ijklm}J_{ij}J_{ik}J_{jl} J_{km}\phi_l\phi_m \\
\nonumber && +\frac{\beta^3}{4}\sum_{ijklm}J_{ij}J_{ik}
J_{il}J_{im}\phi_j\phi_k\phi_l\phi_m
-\beta^4\sum_{injklm}J_{in}J_{ij}J_{ik}
J_{il}J_{nm}\phi_j\phi_k\phi_l\phi_m \\ \nonumber &&
+\frac{\beta^5}{3}\sum_{inpjklm}J_{in}J_{ip}J_{nj}
J_{nk}J_{nl}J_{pm}\phi_j\phi_k\phi_l\phi_m
+\frac{\beta^5}{2}\sum_{inpjklm}J_{in}J_{ip}J_{ij}
J_{ik}J_{nl}J_{pm}\phi_j\phi_k\phi_l\phi_m  \\ \nonumber &&
-\frac{\beta^5}{3}\sum_{injklm}J^2_{in}J_{ij}
J_{ik}J_{il}J_{im}\phi_j\phi_k\phi_l\phi_m
-\frac{\beta^5}{4}\sum_{injklm}J^2_{in}J_{ij}
J_{ik}J_{il}J_{im}\phi_j\phi_k\phi_l\phi_m.
\end{eqnarray}
\end{widetext}
The result (\ref{eq32}) shows that the indirect two-point
interactions of type $JJ\phi\phi$ available in the effective
Hamiltonian (\ref{eq29}) do not exist in this higher accuracy of
the theory. The interactions of type $JJJ\phi\phi$ and
$JJJJ\phi\phi$ presented in the effective Hamiltonian (\ref{eq32})
extend up to distances $3R_{int}$. The same is valid for the four
point interactions included in (\ref{eq32}).

 Within  LWLA the Eq.~(\ref{eq32}) yields
\begin{widetext}
\begin{equation}
\label{eq33}
 {\cal{H}} =  \frac{1}{2}\sum_{\mbox{\boldmath$k$}}\left(r + ck^2
\right)|\phi({\mbox{\boldmath$k$}})|^2   + \frac{u}{N}
\sum_{(\mbox{\boldmath$k$}_1,\mbox{\boldmath$k$}_2,\mbox{\boldmath$k$}_3)}
\phi(\mbox{\boldmath$k$}_1)
\phi(\mbox{\boldmath$k$}_2)\phi(\mbox{\boldmath$k$}_3)
\phi(-\mbox{\boldmath$k$}_1-\mbox{\boldmath$k$}_2-\mbox{\boldmath$k$}_3)\:.
\end{equation}
\end{widetext}
In (\ref{eq33}),
\begin{equation}
\label{eq34} c=\left(1+\frac{5}{z}\right)c_0, \;\;\;
 r=\left(1+\frac{3}{z}\right)\tilde{r}_0,
 \;\;\;u =\left(1+\frac{4}{z}\right)u_0\:,
\end{equation}
where $\tilde{r}_0 = k_B(T - T_c)$ is given by the ``true''
(renormalized) critical temperature
\begin{equation}
\label{eq35} T_c =T_{c0}\left(1-\frac{1}{z}\right).
\end{equation}

In deriving this lattice version of the effective Hamiltonian we
have performed the lattice summations in (\ref{eq32}) in the
reciprocal ($\mbox{\boldmath$k$})$-space with the help of  LWLA:
$J(k) \approx (J-c_0k^2)$.

The present results demonstrate a type of renormalization of the
GL parameters ($c_0$, $r_0$, $u_0$) of the effective Hamiltonian
due to $1/z$-corrections. By a suitable choice of units, one of
these parameters can be kept invariant, for example, equal to
unity. Therefore, within a suitable normalization of the theory,
the field $\phi_(\mbox{\boldmath$x$})$ acquires a $1/z$-correction
as well~\cite{Uzunov:1996}.

The term ${\cal{H}}_f^{(3)}$ in (\ref{eq27}) has the form
\begin{widetext}
\begin{eqnarray}
\label{eq36} {\cal{H}}_f^{(3)}& = &
-\frac{\beta^2}{3}\sum_{ij}J^3_{ij}\frac{\mbox{th}(\beta
a_i)\mbox{th}(\beta a_j)}{\mbox{ch}^2(\beta a_i) \mbox{ch}^2(\beta
a_j)}  -
\frac{\beta^2}{6}\sum_{ijl}\frac{J_{ij}J_{il}J_{jl}}{\mbox{ch}^2(\beta
a_i) \mbox{ch}^2(\beta a_j)\mbox{ch}^2(\beta a_l)} \\ \nonumber &&
-\beta^2\sum_{ijl}J^2_{ij}J_{jl}\frac{\Delta_l\mbox{th}(\beta
a_j)}{\mbox{ch}^2(\beta a_i) \mbox{ch}^2(\beta a_j)}
-\frac{\beta^2}{2}\sum_{ijln}J_{ij}J_{il}J_{jn}\frac{\Delta_l\Delta_n}{\mbox{ch}^2(\beta
a_i) \mbox{ch}^2(\beta a_j)} \\ \nonumber &&
-\frac{\beta^2}{3}\sum_{ijln}J_{ij}J_{il}J_{in}\frac{\Delta_j\Delta_l\Delta_n
\mbox{th}(\beta a_i)}{\mbox{ch}^2(\beta a_i)}.
\end{eqnarray}
\end{widetext}
Let us consider the contribution of the term ${\cal{H}}_f^{(3)}$
to the quadratic ($\phi^2$-) part of the effective Hamiltonian
(\ref{eq33}). In performing the calculations for both short-range
($nn$) and long-range ($R_{int} \gg a_0$) we have  to evaluate
again several lattice sums. Here we shall mention a particular
sum, namely,
\begin{equation}
\label{eq37}
 \frac{1}{N}\sum_{ijl}J_{ij}J_{il}J_{jl}.
\end{equation}
which is equal to zero for $nn$ interactions but gives a
contribution $(\approx J^3/z)$ for interaction radius $R_{int} >
a_0$. Thus we introduce the
 following interpolation formula:  $E = \kappa(z)/z$, where $ 0 \leq \kappa (z) \leq 1$
is an interpolation parameter which is supposed to be a smooth
function of the coordination number $z$. The limiting case $\kappa
=0$ corresponds to $nn$
 interactions and the limiting case $\kappa =1$ corresponds to interactions of
 larger size. We suppose that the shape of the function $\kappa (z)$
depends on details of the function $J(R)$.

Bearing in mind these notes we have calculated the quadratic
($\phi_i\phi_j-$)contribution to the effective Hamiltonian
${\cal{H}}(\phi)$ which comes from the Hamiltonian part
${\cal{H}}^{(3)}_f(\phi)$ given by (\ref{eq36}). Here we present
the results for the parameters $T_{c0}$, $c$, and $r$ which define
the quadratic part ${\cal{H}}_2$ of ${\cal{H}}$:
\begin{equation}
\label{eq38} {\cal{H}}_2(\phi) =
\frac{1}{2}\sum_{\mbox{\boldmath$k$}}
G_0^{-1}(k)|\phi(\mbox{\boldmath$k$})|^2,
\end{equation}

with the (bare) correlation function
\begin{equation}
\label{eq839} G_0^{-1}(k) = {\cal{D}}_1(z)\left[T-T_c +
{\cal{D}}_2(z)T_c\rho^2k^2\right]\:.
\end{equation}
Here
\begin{equation}
\label{eq40} T_c  = \left( 1 - \frac{1+\kappa}{z} -
\frac{1+3\kappa}{3z^2}\right) T_{c0}\:,
\end{equation}
\begin{equation}
\label{eq41} {\cal{D}}_1 (z) = 1 + \frac{3+4\kappa}{z} +
\frac{22+60\kappa + 30\kappa^2}{3z^2}\:,
\end{equation}
and
\begin{equation}
\label{eq42} {\cal{D}}_2 (z) = 1 + \frac{2+3\kappa}{z} +
\frac{14+30\kappa +9\kappa^2}{3z^2}.
\end{equation}
The result (\ref{eq35}) for $T_{c}(z)$ has been published for the
first time~\cite{Uzunov1:1996} in case of $nn$ interactions
($\kappa = 0$); note, that there are errors in
Ref.~\cite{Uzunov1:1996} for the functions ${\cal{D}}_1(z)$ and
${\cal{D}}_2(z)$.

The functions ${\cal{D}}_1(z)$ and ${\cal{D}}_2(z)$ renormalize
the field $\phi(\mbox{\boldmath$k$})$ and the vertices $c_0$, and
$r_0$. For a total renormalization of the parameters of the theory
up to the second order in the $1/z-$expansion we need to know the
$(1/z)^2$-correction to the vertex $u_0$. We suppose that the
calculation of this correction can be accomplished in the way
described above; this ``$z$-renormalization'' has been discussed
to first order in $(1/z)$ in Ref.~\cite{Uzunov:1996}. Here we wish
to stress that within our extension of the theory the magnetic
susceptibility $G_0(0)$ is ${\cal{D}}_1$ times smaller than the
known MF susceptibility corresponding to ${\cal{D}}_1(\infty) =
1$.

The numerical coefficients in (\ref{eq40}) - (\ref{eq42}) indicate
that real numbers $z$ of $nn$ like $n=2,4,6$ for simple lattices
of spatial dimensionalities $D=1,2,3$, respectively, give a good
expansion parameter $1/z$. The $1/z$-orrections are more
substantial for the case of short-range interactions ($R_{int}
\sim a_0$), and one may suppose that for such interactions the
$(1/z)$-series (\ref{eq40}) - (\ref{eq42}) are asymptotic; for the
case
  of $T_c$, see a discussion of this topic in Ref.~\cite{Fisher:1964}. But
  even in case of asymptotic type of these series they may give more reliable
  results than the ``bare'' values
 ($c_0,r_0,u_0$) of the Landau parameters;
  see arguments presented in Ref.~\cite{Fisher:1964}.

\subsection{Critical temperature}

The critical temperature $T_c$ given by (\ref{eq40}) can be
compared with exact and reliable numerical (MC) results. Let us
consider $nn$ interactions ($\kappa = 0$). For one-dimensional
($1D$) IM, we know that $T_c = 0$, MFA predicts $T_c = 2J_0/k_B$
(in this case, $z =2D= 2$), and (\ref{eq40}) yields $T_c=
5J_0/6k_B$. This is a quite good result for $1D$ systems with very
strong fluctuation effects. In $2D$ systems the fluctuations are
not so strong and we find that (\ref{eq40}) reproduces the exact
Onsager result ($T_c = 2.27 J_0/k_B$) with an error of 22\%, i.e.,
we have $T_c = 35J_0/12k_B$. For $3D$ systems our result is $T_c =
89J_0/18k_B$, whereas the best series analysis and MC results
yield a difference of 9\%: $T_c = 4.5J_0/k_B$ (see, e.g.,
Refs.\cite{Baker:1994, Baker:1996}). Our results seem quite
reliable. Let us emphasize, that the $1/z$-series, as almost all
most relevant series, known in theoretical physics, is an
asymptotic series. Therefore, one may expect, that the results for
$T_c$ will be worsened after some order in ${1/z}$, for example,
fourth order for $T_c(3D)$, and third, or, even second order for
$T_c(2D)$.

\subsection{Ground state} The $1/z^2$-correction to the vertex
$u_0$ has not been yet calculated although this calculation does
not present difficulties. In this situation we shall give notion
for the ground state energy by using the first order corrections
to $r_0$ and $u_0$.  The equilibrium free energy per site $f =
(F/N)$ is given by $F={\cal{H}}$ and (\ref{eq33}) for the
$(k=0)$-Fourier amplitude $\phi(0)$, which minimizes $f$. Denote
for convenience $\phi(0) = \sqrt{N}\varphi \neq 0$ for the
low-temperature ordered phase. From (33) we obtain $f = -(r^2/16u)
< 0$ whereas the ``unrenormalized free energy is $f_0 =
({\cal{H}}_0 /N) = -(r_0^2/16u_0) < 0$ Thus, using (\ref{eq34}) we
have  $f(T) =(1+2/z)f_0(T)$, which means that the new effective
theory has a lower energy of the ordered phase. This is true also
in the case of $T=0$, where $r_0 = -k_BT_{c0} = -J$. For the zero
temperature (ground) state we have $f_0(0) =-3J/4$ and  $f(0) = -3
(1 + 2/z)J/4$. This result is also along the right direction
because the MF theories (${\cal{H}}_0$) give unreliable high
values of the ground state energy. The order parameter
$\varphi^2(T) = (-r/4u) $ is $(1-\/z)$ times smaller than the
respective quantity $\varphi^2_0(T) = (-r_0/4u_0) $ in the usual
theory based on ${\cal{H}}_0$.

\subsection{Effective interactions and growing of fluctuation
correlations}

We have explicitly shown the phenomenon of the growing length size
of the interpaticle correlations in a classic system of
interacting particles. To see this we have already introduced a
new interpretation of the terms in the effective Hamiltonian (see
Sec.II.D). Let us consider the terms present in ${\cal{H}}$ as
terms describing certain inter-site interactions. While the
initial interaction $J_{ij}$ in IM ensures only two-site
correlations (interactions), the effective Hamiltonians
(\ref{eq15}), (\ref{eq29}), and (\ref{eq32}) contain multi-site
effective interactions. In contrast to the usual theory
(\ref{eq15}), where only extremely short-range effective
correlations are contained, the more precise effective
Hamiltonians contain long-range two-site ($\phi_i\phi_j$) and
four-site ($\phi_i\phi_j\phi_k\phi_l$) correlations, and all of
these correlations are indirect, i.e. the correlation, for
example, between two sites $(ij)$ is mediated by one or more other
sites $(k,...)$. A direct $(J_{ij})$-interaction is presented by
the first term in the r.h.s. of (\ref{eq15}) but also the system
exhibits two indirect correlations of type $\phi_i\phi_j$ and
$\phi_i\phi_j\phi_k\phi_l$ given by the last two terms in the
r.h.s. of (\ref{eq15}). In the more precise variants of the
theory, where a larger portion of the initial partition function
has been calculated, the direct inter-site interaction vanishes,
and the particles are correlated only by indirect effective
interactions. The length scale of these correlations grows in a
monotonous way
 with the increase of the accuracy of the calculation,
i.e. with the increase of the number $l$ of the terms in the
series (\ref{eq27}). If we take the two-site correlations in the
$nn$ IM as an example, the maximal length of extension of these
correlations in (\ref{eq1}) is $a_0$, in (\ref{eq15}) -- $2a_0$,
in (\ref{eq29}) -- $3a_0$, in (\ref{eq32}) --  $4a_0$, i.e.
$(p-1)a_0$, where $p$ is the maximal number of the summation
indices in the terms of type $\phi_i\phi_j$ in a given effective
Hamiltonian. Surely, the number $p$ tends to $N$. This means that
the most accurate effective field theory of many-body systems will
correspond to (almost-) infinite range of correlations.

The origin of these correlations is purely statistical. This
effect is known and has both general formulation and application
in many-body physics. Here we have established and described in
details the concrete mechanism of this effect and, moreover, we
have performed a demonstration of the remarkable picture of
successive growth of the correlation length scale.

\subsection{Final remarks}

Obviously the $(1/z)$-corrections are not the main point of
discussion at the end of this paper of restricted length. Let us
mention that the growth of the correlations discussed in Sec.
III.F is not related with the $1/z$-corrections. It exists for any
$z$, even in the ``MF limiting case'' of $z \rightarrow \infty$,
when the GL parameters keep their ``initial'' values
$(c_0,r_0,u_0)$. This effect follows from the fact that the terms
in the initial Hamiltonian are compensated by the first
``fluctuation'' correction; see the first term in the r.h.s. of
(\ref{eq27}). At the next level of accuracy of the calculation,
terms coming from the $(l=2)$-term in (\ref{eq27}) compensate the
available terms and this process continues up to the incorporation
of all particles in the correlation phenomenon; remember that the
term corresponding to $l=1$  is equal to zero.

Here we emphasize that the terms in ${\cal{H}}_0$ -- actually one
of the most often used Hamiltonian, does not exist at all. They
vanish just after the inclusion of the first correction to the
usual theory; see $\Delta_i\Delta_j$-correction in (\ref{eq27}).
In place of these terms other terms with more complex structure
come from the perturbation series (\ref{eq27}). The outlined
picture clearly indicates, that the terms which finally remain in
the $\phi^4$-theory, are terms of type
\begin{equation}
\label{eq43} \frac{1}{2}\left(\beta^{M-1}J^{M} -
\beta^{M}J^{M+1}\right)\phi^2, \;\;\;\;\; M \sim N\:,
\end{equation}
where obvious notations have been introduced; for example,
$\beta^1J^{2}$ denotes the first term in the r.h.s. of
(\ref{eq29}). The $\phi^4$ terms behave differently, because a
lowest order term in $J$,namely, a term of type $\beta^3
J^4\phi^4$ appear at any step of development of the series
(\ref{eq27}).

At any stage of this surprising picture of the infinite series of
successive modifications of the Hamiltonian both $\phi^2$- and
$\phi^4$-terms keep their numerical coefficients equal to that in
the usual GL Hamiltonian ${\cal{H}}_0$. This is true within the
whole scope of validity of the expansion in powers of $\phi$.

An important note, which should be emphasized is the following.
While the sum~(\ref{eq2}) is invariant with respect to the site
$i$ in regular lattices, the sum~(\ref{eq7}) depends on the site
$i$. The reason is that the field configuration $\{\phi_i\}$ which
takes part in Eq.(\ref{eq7}) is not the equilibrium field. For the
equilibrium field $\bar{\phi}_i$ the sum~(\ref{eq7}) will not
dependent on the site $i$. This is consistent with the general
notion that the equilibrium order in the volume of a homogeneous
system in lack of effects of external fields, should be uniform.

Our consideration justifies the GL fluctuation Hamiltonian.
However, we have presented a new and quite surprising picture of
the inter-particle correlations, which reveals new remarkable
properties of the GL theory. Apart from the $1/z$-corrections to
the GL parameters, the structure of this theory is absolutely
comprehensive as a tool for investigation of large-scale
correlation phenomena in many-body systems. We are certain that
our findings have an application beyond the field of phase
transitions.\\

{\bf ACKNOWLEDGEMENTS:}

 The author thanks the hospitality at
Department of Physics and Astronomy, University of Western
Ontario, London, Canada. The research grant Ph. 1507 (NFSR-Sofia)
is also acknowledged.


\begin{thebibliography}{11}
\bibitem{Note:2008} Remember that the thermodynamic states
 represented by the variable (``lattice field") $\phi$ in
(5) correspond to stable, metastable or (absolutely) unstable
thermodynamic equilibria, $\phi_0$, given by the variation
condition $\delta {\cal{H}}(\phi_0) = 0$. In this interpretation
of Eq.~(5) both equilibrium ($\phi_0$) and nonequilibrium  ($\phi
\neq 0$) states can participate in the function ${\cal{H}}(\phi)$.
For this reason, the latter can be referred to as
``non-equilibrium free energy" with equilibrium part, given by
${\cal{H}}(\phi_0)$.
\end{thebibliography}
\end{document}